\documentclass[3p,twocolumn]{elsarticle}

\usepackage{hyperref}
\usepackage{amsmath,amssymb}

\journal{Journal of \LaTeX\ Templates}

\usepackage{xcolor}

\begin{document}

\begin{frontmatter}

\title{Enhanced pion-to-proton ratio at the onset of the QCD phase transition}


\author[coe]{Thiranat Bumnedpan}
\author[fias]{Jan Steinheimer}
\author[itp,hfhf,gsi]{Marcus Bleicher}
\author[coe]{Ayut Limphirat}
\author[coe]{Christoph Herold\corref{mycorrespondingauthor}}
\cortext[mycorrespondingauthor]{Corresponding author}
\ead{herold@g.sut.ac.th}

\address[coe]{Center of Excellence in High Energy Physics \& Astrophysics, Suranaree University of Technology, Nakhon Ratchasima, 30000, Thailand}
\address[fias]{Frankfurt Institute for Advanced Studies, Ruth-Moufang-Str. 1, 60438 Frankfurt am Main, Germany}
\address[itp]{Institut f\"{u}r Theoretische Physik, Goethe Universit\"{a}t Frankfurt, Max-von-Laue-Strasse 1, Frankfurt am Main, 60438, Germany}
\address[hfhf]{Helmholtz Research Academy Hesse for FAIR (HFHF), GSI Helmholtz Center for Heavy Ion Physics, Campus Frankfurt, Max-von-Laue-Str. 12, Frankfurt, 60438, Germany}
\address[gsi]{GSI Helmholtzzentrum f\"{u}r Schwerionenforschung GmbH, Planckstr. 1, Darmstadt, 64291, Germany}

\begin{abstract}
The pion-to-proton ratio is identified as a potential signal for a non-equilibrium first-order chiral phase transition in heavy-ion collisions, as the pion multiplicity is directly related to entropy production. To showcase this effect, a non-equilibrium Bjorken expansion starting from realistic initial conditions along a Taub adiabat is used to simulate the entropy production. Different dynamical criteria to determine the final entropy-per-baryon number are investigated and matched to a hadron resonance gas model along the chemical freeze out curve to obtain the final pion and proton numbers. We detect a strong enhancement of their multiplicity ratio at the energies where the system experiences a strong phase transition as compared to a smooth crossover which shows almost no enhancement. 
\end{abstract}

\begin{keyword}
\texttt{elsarticle.cls}\sep \LaTeX\sep Elsevier \sep template
\MSC[2010] 00-01\sep  99-00
\end{keyword}

\end{frontmatter}


\section{Introduction}

A quark-gluon plasma is created in heavy-ion collisions at sufficiently large center-of-mass energies \cite{Heinz:2000bk,STAR:2005gfr}. It is a new phase of strongly interacting matter that is characterized by deconfinement and the restoration of chiral symmetry. Although lattice QCD has ruled out the possibility of a quantum chromodynamics (QCD) phase transition for small values of the baryochemical potential $\mu_{\rm B}$, a critical end point (CEP) and first-order phase transition (FOPT) are possible to be present in the QCD phase diagram at high baryon densities. The location of the CEP has been estimated from first principles \cite{Fischer:2014ata,Gao:2020fbl} and also effective models \cite{Scavenius:2000qd,Schaefer:2004en,Fukushima:2008wg} with so far only little agreement between the various theoretical approaches. 

Experimentally, the focus on investigating the QCD phase diagram has been on the measurement of cumulants or cumulant ratios of baryon number, strangeness, electric charge, e.g. in the beam energy scan program at STAR \cite{STAR:2021iop}, at NA49/61 \cite{Grebieszkow:2009jr,Andronov:2018ccl}, and for lowest energies, also at HADES \cite{HADES:2020wpc}. It is expected that the presence of a CEP will manifest in non-monotonic behavior of these observables as function of beam energy, as shown in lattice QCD \cite{Karsch:2016yzt,Bazavov:2020bjn}, functional techniques \cite{Isserstedt:2019pgx}, and effective models \cite{Skokov:2010uh,Almasi:2017bhq,Wen:2018nkn}. Due to the sensitivity of these cumulants to the correlation length, it is not fully understood how much of the signal survives the nonequilibrium dynamics of the expanding medium, finite size and time effects \cite{Berdnikov:1999ph}, hadronic rescattering, as explored by a multitude of dynamical models in recent years \cite{Athanasiou:2010kw,Stephanov:2011pb,Nahrgang:2011mg,Mukherjee:2015swa,Jiang:2015hri,Herold:2016uvv,Herold:2017day,Stephanov:2017ghc,Stephanov:2017wlw,Nahrgang:2018afz,Nahrgang:2020yxm,Du:2020bxp}. 

This possible caveat of the impact of nonequilibrium dynamics on observables can be turned into an advantage when considering signals for a FOPT. Spinodal decomposition and the emergence of density inhomogeneities have been widely studied \cite{Randrup:2009gp,Randrup:2010ax,Steinheimer:2012gc,Herold:2013qda,Herold:2014zoa,Jiang:2017fas,Poberezhnyuk:2020cen}. Besides that, the FOPT sees a delayed relaxation of the critical mode which consequently produces additional entropy \cite{Csernai:1992as}, an effect that has been shown to potentially double the initial entropy-per-baryon number ($S/A$) \cite{Herold:2018ptm}. 

In the present letter, we intend to show the significant effect that this additional entropy can have on a simple observable such as the pion to proton multiplicity ratio, which can be measured much more easily than higher order cumulants or correlation functions.
To do so we will start from the relaxational dynamics of the chiral order parameter which is coupled to a Bjorken expansion. Realistic initial conditions are obtained for low- to intermediate energies from the stationary one-dimensional Taub adiabat model \cite{Taub:1948zz,Thorne:1973}. We observe differences between an ideal hydrodynamic evolution and the full nonequilibrium dynamics near the CEP and across the FOPT. After comparing different criteria for obtaining the final $S/A$, we use two parametrized freeze-out curves to map this value to pion and proton multiplicities via a hadron resonance gas model \cite{Vovchenko:2019pjl}. 

After a description of the model in Section~\ref{sec:model}, we report our results in Section \ref{sec:results}, and finally conclude with a summary in Section \ref{sec:summary}.

\section{Model description}
\label{sec:model}

To describe the entropy production due to a non-equilibrium first order chiral phase transition the chiral Bjorken expansion, introduced in \cite{Herold:2018ptm} as a simple variant of non-equilibrium chiral fluid dynamics \cite{Nahrgang:2011mg}, is employed. This simple 1D model yields a dynamics that is very similar to the event-averages of more sophisticated (3+1)D models \cite{Herold:2013qda}. As shown in \cite{Herold:2018ptm}, the model produces entropy across a FOPT due to the relaxational dynamics of the non-equilibrium order parameter evolution, obeying a Langevin equation of motion, 
\begin{equation}
 \label{eq:eom_sigma}
 \ddot\sigma+\left(\frac{D}{\tau}+\eta\right)\dot\sigma+\frac{\delta\Omega}{\delta\sigma}=\xi~.
\end{equation}
Here, the dots above $\sigma$ denote derivatives with respect to proper time $\tau$. Furthermore, $\Omega$ is the mean-field grand canonical potential and we set $D=1$ in the Hubble term for a longitudinal expansion. The $T$- and $\mu$-dependent damping coefficient $\eta$ is related to the stochastic noise $\xi$ with the dissipation-fluctuation relation 
\begin{equation}
\label{eq:dissfluct}
 \langle\xi(t)\xi(t')\rangle=\frac{m_{\sigma}\eta}{V}\coth{\left(\frac{m_{\sigma}}{2T}\right)}\delta(t-t')~,
\end{equation} 
with the fireball volume $V$, and the sigma screening mass $m_{\sigma}$ which is given by the second derivative of $\Omega$ with respect to $\sigma$ at the equilibrium state. The function $\eta(T,\mu)$ has been derived from the 2PI effective action and vanishes around the CEP to allow for the emergence of long-range fluctuations. Physically, it arises from various sigma-meson scatterings, see \cite{Herold:2018ptm} for further details. Alternatively, simplifications with a constant damping coefficient have been studied earlier \cite{Biro:1997va}. Note that the choice of $\eta$ will directly impact the relaxation time of the system.

In this study, the mean-field potential $\Omega$ is obtained from a quark-meson (QM) model, as integral of the quark degrees of freedom, 
\begin{align}
\label{eq:Lagrangian}
 {\cal L}&=\overline{q}\left(i \gamma^\mu \partial_\mu-g \sigma\right)q + \frac{1}{2}\left(\partial_\mu\sigma\right)^2- U(\sigma)~, \\
 U(\sigma)&=\frac{\lambda^2}{4}\left(\sigma^2-f_{\pi}^2\right)^2-f_{\pi}m_{\pi}^2\sigma +U_0~,    
\end{align}
where we use light quarks $q=(u,d)$ only and standard parameters $f_\pi=93$~MeV, $m_\pi=138$~MeV, $\lambda^2=19.7$. The QM model is far from a realistic description of dense QCD matter. However, it contains the relevant symmetries and necessary degrees of freedom to describe a chiral transition at finite baryon density and it allows us to solve the proper equations of motion for the chiral field. 

To relate the trajectories which can be calculated with the chiral Bjorken expansion to a beam energy the corresponding initial state, i.e. initial temperature $T$ and density as well as entropy per baryon needs to be calculated self consistently from the equation of state. This can be done reliably assuming that most entropy in the initial state of low energy heavy ion collisions is produced by shock heating. Then, the initial compression can be estimated by a one-dimensional shock wave solution, the Rankine-Hugoniot-Taub adiabat \cite{Taub:1948zz,Thorne:1973}. Then, one has to solve the following equation \cite{Motornenko:2019arp}:
\begin{equation}
    \label{eq:Taub}
    (P_0+\varepsilon_0)\, (P+\varepsilon_0)\, n^2=(P_0+\varepsilon)\, (P+\varepsilon)\, n^2_0\,,
\end{equation}
where $P_0=0$, $\varepsilon_0/n_0 - m_N=-16$ MeV and $n_0=0.16~ {\rm fm^{-3}}$ are the ground state pressure, energy density, and baryon density. With any known relation $P=P(\varepsilon, n)$, Eq.~\ref{eq:Taub} can be solved. Furthermore, the collision energy can be related to the compression as:
\begin{equation}
    \label{eq:stopping}
    \gamma^{\rm CM}=\frac{\varepsilon n_0}{\varepsilon_0 n},~\gamma^{\rm CM}=\sqrt{\frac{1}{2}\left(1+\frac{E_{\rm lab}}{m_N}\right)}\,.
\end{equation}
Here $\gamma^{\rm CM}$ is the Lorentz gamma factor in the center of mass frame of the heavy ion collisions and $E_{\rm lab}$ is the beam energy per nucleon in the laboratory frame of a fixed target collision.

\section{Results}
\label{sec:results}

\begin{figure}[tbp]
\centering
\includegraphics[width=.45\textwidth,origin=c]{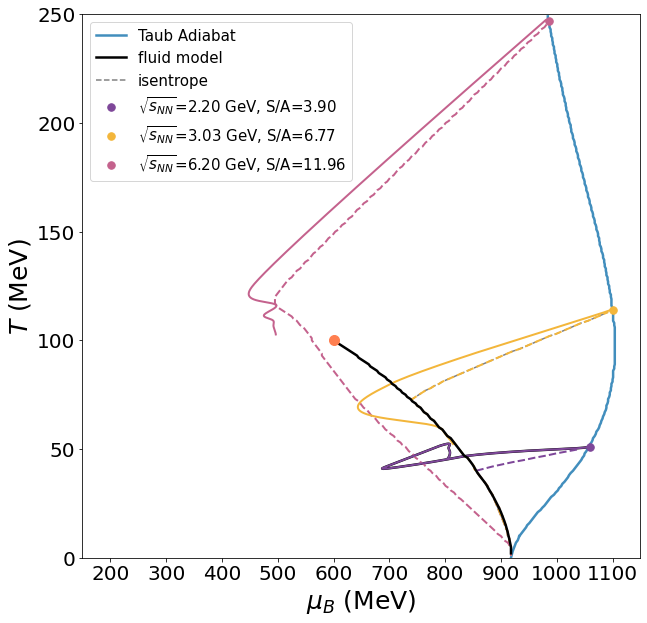}
\caption{Evolutions of the fluid dynamical model (solid lines) compared to the isentropes (dashed line) from the same initial condition along the Taub adiabat. The position of the CEP and FOPT are indicated by the dot and solid black line.}
\label{fig:traj}
\end{figure}

Fig.~\ref{fig:traj} depicts trajectories for initial conditions at three different beam energies, defined by values of $S/A$. The initial conditions are shown as dots along the blue Taub adiabat line. The solid black curve and adjacent dot around $(T,\mu_{\rm B})$ show the location of the FOPT and CEP of the quark-meson model in mean-field. For each initial condition the non-equilibrium evolution (solid lines), averaged over $10^4$ chiral Bjorken dynamics events, are compared to isentropes (dashed lines) along which $S/A$ remains constant and which represent an ideal hydrodynamic evolution. We use these as proxy for the evolution without phase transition, where the relaxational dynamics plays only an insignificant role and the produced entropy is marginal compared to a scenario with a FOPT. While the trajectories halt at a final time of $\tau=12$~fm from our simulation, the isentropes are drawn until the density reaches that of ground state nuclear matter. We notice several differences between nonequilibrium and ideal hydrodynamics, most notable is the overshooting of the FOPT line at lowest energies which is not present in the isentropes that stay along the phase transition line and then lie slightly below it. As a result, the freeze-out points for these two scenarios will be different. 

\begin{figure}[tbp]
\centering
\includegraphics[width=.45\textwidth,origin=c]{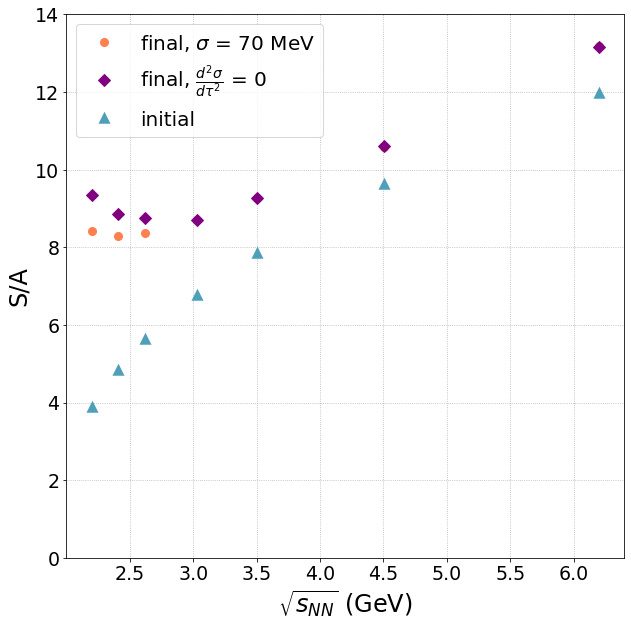}
\caption{Initial and final entropy-per-baryon number, the latter one extracted for two different freeze-out conditions. In both cases, a clear enhancement at low energies is found compared to the initial value. }
\label{fig:spa}
\end{figure}

For defining a freeze-out criterion, we compare two different scenarios: Since the model focuses on the sigma field, we use its dynamics to define a point where the entropy production ends: First, at a constant value which we set to $\sigma_{\rm f.o.}=70$~MeV, and second, at the point where the slope in the function $\sigma(\tau)$ remains constant, indicating the completion of the transition, i.e. $\mathrm d^2\sigma/\mathrm d \tau^2\vert_{\sigma=\sigma_{\rm f.o.}}=0$. Fig.~\ref{fig:spa} shows the final $S/A$ for both cases as function of $\sqrt{s_{NN}}$ together with the initial $S/A$ from the Taub-adiabat. We see that, while the initial $S/A$ increases monotonically with energy, both of the final ones are strongly enhanced at lowest $\sqrt{s_{NN}}$. For $\sqrt{s_{NN}}\leq 3.0$~GeV, both yield nearly the same entropy-per-baryon number of the final state, below that, the discrepancy is of the order of $\sim 10$\% only. Since both freeze-out conditions yield similar results, from here on, we will discuss only the criterion of the vanishing second derivative which produces a slightly stronger effect. 

\begin{figure}[tbp]
\centering
\includegraphics[width=.45\textwidth,origin=c]{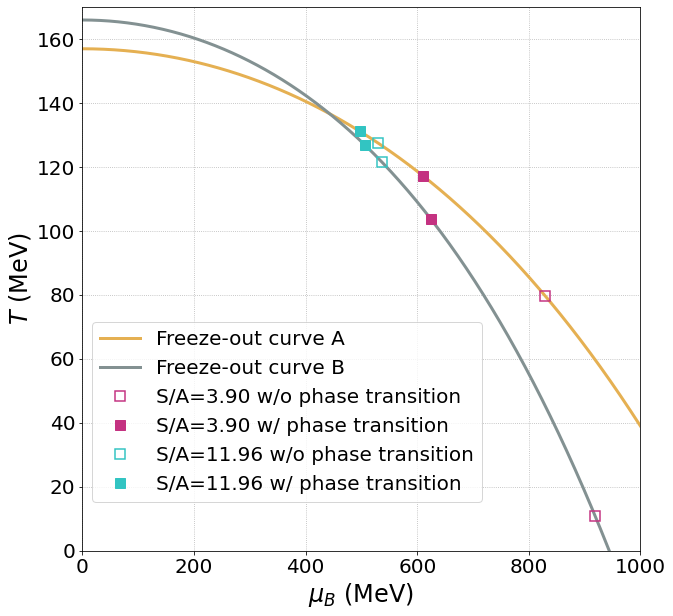}\caption{ Different parametrizations of the freeze-out curve with the freeze-out points of the evolutions with the smallest/largest initial $S/A$ as well as with and without a phase transition.}
\label{fig:focurves}
\end{figure}

To translate our final state entropy into particle multiplicities, we match the obtained values of $S/A$ onto a parametrized freeze-out curve by identifying $T$ and $\mu_{\rm B}$ along that curve with the final $S/A$. After that, we use these values of $T$ and $\mu_{\rm B}$ to calculate the pion and proton numbers from a hadron resonance gas model, using the Thermal-FIST (Thermal, Fast and Interactive Statistical Toolkit) package \cite{Vovchenko:2019pjl}. As the final entropy in our approach is directly connected to a beam energy we can then investigate the beam energy dependence of the particle multiplicities. Since there is currently no agreement on the precise location of the freeze-out curve, especially for the lowest energies, we compare results from two versions obtained from thermal model fits to experimental data over a wide range of energies using the parametrization
\begin{equation}
    \label{eq:freeze}
    T_{\rm f.o.}(\mu_B) = a - b\mu_B^2 - c\mu_B^4~.
\end{equation}
with parameter set A as $a=0.157$~GeV, $b=0.087$~GeV$^{-1}$, and  $c=0.092$~GeV$^{-3}$ (freeze-out curve A \cite{Vovchenko:2015idt}) and parameter set B as $a=0.166$~GeV, $b=0.139$~GeV$^{-1}$, and  $c=0.053$~GeV$^{-3}$ (freeze-out curve B \cite{Cleymans:2005xv}). 
These freeze-out curves, together with the freeze-out points, with and without a phase transition, corresponding to the evolutions with the smallest and largest beam energy, are shown in Fig.~\ref{fig:focurves}. While the freeze-out coordinates lie relatively close together for an initial $S/A$ of $11.96$, they significantly differ for the smallest initial $S/A$ of $3.9$, especially the two points with phase transition lie far away from each other. 

\begin{figure}[tbp]
\centering
\includegraphics[width=.45\textwidth,origin=c]{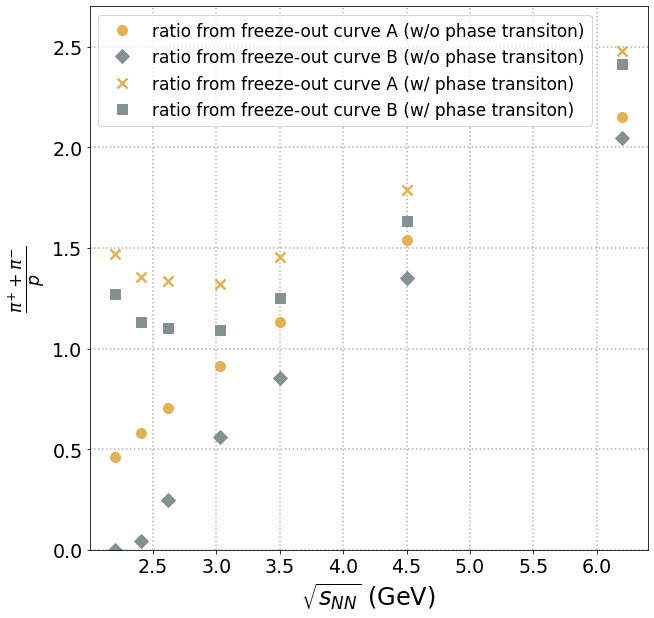}
\caption{Pion-to-proton multiplicity ratio as function of center-of-mass energy for scenarios with and without phase transition. For both freeze-out curves, we see a strong enhancement at low energies for scenarios with a phase transition. 
}
\label{fig:pitop}
\end{figure}

This difference is reflected in the ratio of charged pion to proton numbers obtained from the HRG model, see figure~\ref{fig:pitop}. Here, again, we show results for freeze-out curves A and B, each with and without phase transition. In general, we observe an enhancement of the multiplicity ratio in the presence of a phase transition (cross and square symbol) compared to a no-transition scenario (circle and diamond symbol) which is most significant at the low values of $\sqrt{s_{NN}}$, where the values go up again after an initial decrease from low energies. Since freeze-out curve A lies above curve B throughout the investigated energy range, the same is true for the corresponding pion-to-proton ratios. We see that as we approach the lowest center-of-mass energies around $2.2$~GeV, the values approach zero for curve B without phase transition. For curve B, a ratio of $0.45$ is the lowest one, which is clearly more realistic. Qualitatively, however, both curves give a consistent result. 

To understand why the pion to proton ratio keeps increasing for low beam energies, in the scenario with a non-equilibrium phase transition, it is important to emphasize that the Taub adiabat and with it every initial state constructed from the quark-meson Lagrangian lies entirely within the phase of chiral symmetry restoration. In other words, even initial states at the lowest beam energies will have to pass through the chiral phase transition. In a more realistic setup the systems created at the lowest beam energies would rather remain within the hadronic phase from the beginning. For these evolutions, there would be no entropy gain from a nonequilibrium FOPT. Therefore we predict that a beam energy scan around these values of $\sqrt{s_{NN}}$, where the FOPT is reached should then reveal a sudden increase of the pion-to-proton ratio at the collision that passes through that transition.

\section{Summary and conclusions}
\label{sec:summary}

We have studied the influence of a chiral FOPT on the entropy-per-baryon number and pion-to-proton ratio from a Bjorken expansion coupled to the nonequilibrium evolution of the order parameter and a calculation of particle multiplicities along two parametrizations of freeze-out conditions obtained from experimental data. We have found that a nonequilibrium FOPT leads to a significant increase of $S/A$ compared to a case without phase transition and, experimentally, to an enhancement in the multiplicity ratio. This result is qualitatively independent of the specific choice of freeze-out criterion. It is also reasonable to expect that the gain in entropy is a generic feature of the nonequilibrium phase transition rather than the underlying model of the arguably simple quark-meson Lagrangian. If indeed a QCD FOPT is present at large baryochemical potentials, it should be possible to unveil it within a beam energy scan at low $\sqrt{s_{NN}}$. Since recent STAR data shows that the flow at $\sqrt{s_{NN}}=3$ GeV can be best described with a stiff hadronic equation of state \cite{STAR:2021yiu}, a sudden jump in the pion-to-proton number at a higher beam energy would clearly signal the lowest center-of-mass energy where the created fireball is consisting of matter above the chiral phase transition line.

\section*{Acknowledgments}

This research has received funding support from the NSRF via the Program Management Unit for Human Resources \& Institutional Development, Research and Innovation [grant number B16F640076]. J.S. thanks the Samson AG and the BMBF through the ErUM-Data project for funding. This work was supported by a PPP program of the DAAD.

\bibliography{mybib}

\end{document}